\documentclass[12pt,aps,prl,a4paper,titlepage, oneside, english]{book}

\usepackage{lmodern}
\usepackage[T1]{fontenc}

\usepackage[utf8]{inputenc}               % encoding del documento

\usepackage[font=small]{quoting}          % citazioni

\usepackage{epigraph}

\usepackage{indentfirst}                  % indenta il primo paragrafo
\usepackage{mparhack}
\usepackage{fixltx2e}
\usepackage{relsize}

\usepackage{fancyhdr}   %Posizione numerazione pagine e titoletti in cima alle pagine
\newlength\FHoffset
\fancyheadoffset{\FHoffset}

\newlength\FHleft
\newlength\FHright

\newbox\FHline
\setbox\FHline=\hbox{\hsize=\paperwidth%
  \hspace*{\FHleft}%
  \rule{\dimexpr\headwidth-\FHleft-\FHright\relax}{\headrulewidth}\hspace*{\FHright}%
}

\usepackage[pdftex,a4paper]{geometry}     % forma e dimensioni della pagina
\usepackage{setspace}
\newcommand*{\persspacing}{\setstretch{.8}}
\usepackage[font=small,labelfont=bf,labelsep=period,aboveskip=6pt]{caption}
\let\oldcaption\caption
\renewcommand{\caption}{\persspacing\oldcaption}
\usepackage[english]{babel}                % internazionalizzazione
\usepackage{latexsym}                      % simboli strani
\usepackage{amsmath}                       % matematica AMS
\usepackage{amssymb}                       % simboli AMS
\usepackage{amsthm}                        % teoremi
\usepackage{amsfonts}
\usepackage{mparhack}
\usepackage{fixltx2e}
\usepackage{relsize}
\usepackage[utf8]{inputenc}
\usepackage{mathtools}
\usepackage{parskip}
\usepackage{comment}
\usepackage{subfigure}
\usepackage{bm}

\newtheorem*{theorem*}{Theorem}
\newtheorem*{defn}{Definition}

\usepackage{url}

\usepackage{breakurl}
\usepackage[pdftex]{hyperref}              % hypernation
\hypersetup{%
	linktocpage=true,%
	pdfstartpage=1,%
	pdfstartview=FitV,%
	breaklinks=true,%
	pdfpagemode=UseNone,%
	pageanchor=true,%
	pdfpagemode=UseOutlines,%
	plainpages=false,%
	bookmarksnumbered,%
	bookmarksopen=true,%
	bookmarksopenlevel=1,%
	hypertexnames=true,%
	pdfhighlight=/O,%
	urlcolor=blue,%
	linkcolor=blue,%
	citecolor=blue,%
	pdftitle=Tesi,%
	pdfauthor=Bianca Rizzo,%
	pdfsubject=Fisica,%
}
\usepackage{graphicx}                      % inclusione di grafica e immagini
\usepackage[%                              % caption style
	font=small,
	format=hang,
	labelfont=sc,
    textfont=sl,
    labelsep=period
]{caption}
\usepackage{wrapfig}
\usepackage{pdfpages}
\usepackage{enumerate}
\usepackage{multirow}

\usepackage{soul}                          % \hl per l'evidenziazione

\usepackage{lipsum}
\usepackage[colorinlistoftodos]{todonotes}

\usepackage{frontespizio}
\usepackage[linesnumbered,ruled]{algorithm2e}
\usepackage{xparse}

\newsavebox{\fminipagebox}
\NewDocumentEnvironment{fminipage}{m O{\fboxsep}}
 {\par\kern#2\noindent\begin{lrbox}{\fminipagebox}
  \begin{minipage}{#1}\ignorespaces}
 {\end{minipage}\end{lrbox}%
  \makebox[#1]{%
    \kern\dimexpr-\fboxsep-\fboxrule\relax
    \fbox{\usebox{\fminipagebox}}%
    \kern\dimexpr-\fboxsep-\fboxrule\relax
  }\par\kern#2
 }
 
\newcommand\summaryname{Abstract}
\newenvironment{Abstract}%
    {\small\begin{center}%
    \bfseries{\summaryname} \end{center}}

\begin{document}

\begin{figure}
\centering
\includegraphics[scale=0.07]{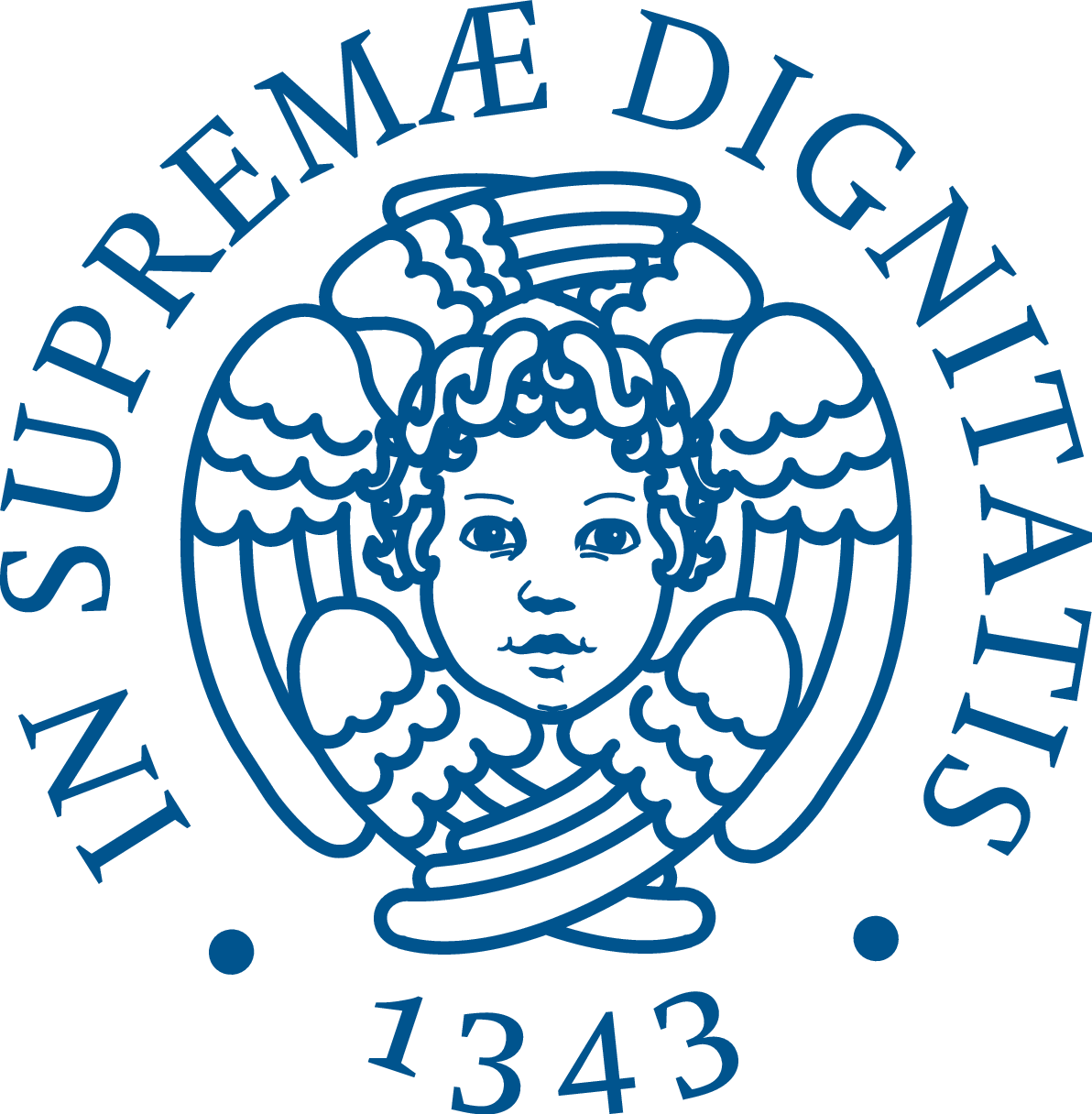}
\end{figure}

\def\blurb{% 

\vspace{24pt} \LARGE \textbf{Universit\`a di Pisa} \\ \vspace{-12pt} \hrulefill \vspace{-12pt} \\ [1em]
  \Large \textsc{Dipartimento di Fisica} }

\def\ligne#1{%
  \hbox to \hsize{%
    \vbox{\centering #1}}}%
\def\haut#1#2#3{%
  \hbox to \hsize{%
    \rlap{\vtop{\raggedright #1}}%
    \hss
    \clap{\vtop{\centering #2}}%
    \hss
    \llap{\vtop{\raggedleft #3}}}}%
\def\bas#1#2#3{%
  \hbox to \hsize{%
    \rlap{\vbox{\raggedright #1}}%
    \hss
    \clap{\vbox{\centering #2}}%
    \hss
    \llap{\vbox{\raggedleft #3}}}}%

\thispagestyle{empty}\vbox to .9\vsize{%
  \vss
  \vbox to 1\vsize{%
    \haut{}{\blurb}{}
    \vfill
    \ligne{\huge \textbf{\textsc{How perturbing a classical 3-spin chain can lead to quantum features}}\\ \vspace{5mm} \Large Tesi di Laurea Triennale (Bachelor thesis)}
    \vspace{5mm}
    %\ligne{\Large Vincent \fsc{Zoonekynd}}
    \vfill
        \vspace{5mm}
    \ligne{%
    \flushleft{
      \large Candidate:\\
      \textbf{Rizzo Bianca}}
      \vspace{-53pt}
      \flushright{
      \large Advisor:\\
      \textbf{Professor Hans-Thomas Elze}\\ }
      }
      \vfill
      \vspace{12pt}
      \hrulefill
      \vspace{12pt}
      \ligne{\large \textbf{Anno Accademico 2019-2020}}
    }%
  \vss
  }

\newpage

\begin{Abstract}
\begin{changemargin}{1cm}{1cm}
It is well known that Quantum Mechanics (QM) describes a world that follows completely counter intuitive rules, that has famous paradoxes as postulates (e.g. superposition of states, wave/particle dualism, the measurement problem...), but that at the same time it is the foundation for the description of the physical world that we experience. While in the Copenhagen Interpretation these paradoxes are fully accepted as a reality, there are other interpretations that try to find a compromise, between the classic macroscopic world and the QM results.\\
A quite interesting one is the Cellular Automata Interpretation of QM, by Gerard 't Hooft \cite{Hooft2014}, according to whom particles evolve following the rules of Cellular Automata (CA), a mathematical model consisting of discrete units that evolve following deterministic laws in discrete space and time. The states of a Cellular Automaton are, by definition, classical and thus deterministic and do not form superpositions: they are identified, in this interpretation, as Ontological States. These states are the ones in which microscopic particles find themselves in reality, whether they are observed or not.
Unfortunately, it is not always possible to determine exactly the Ontological States. To overcome this obstacle, the formalism of QM works as a "\textit{template}" with which a scientist can work and get reliable results even without fully knowing what is happening at the microscopic level.
The problem of measuring these states is solved by considering the observer and his/her instruments as a separate physical system which is equally deterministic, so the act of measurement is interpreted as the interaction between the two systems.

Since it is not possible to know how to demonstrate the underlying classical deterministic structure and dynamics at the smallest microscopic scales at present, what we pursue in this thesis, besides summarizing the concept of the Cellular Automaton Interpretation, is to show that quantum phenomena, in particular superposition states, can arise in a deterministic model because of the limited precision of measurements.   \\
In order to do that we follow the reference \cite{Elze2020}, considering a classical system of three coupled Ising spins. The dynamics is introduced with an operator that permutes the states of the spins: this is interesting also because it is possible to write this operator in terms of Pauli matrices. By doing this, the operator can be applied on the states of the classical system, and at the same time it can be inserted in the context of an Hilbert space and applied on a quantum system as well, such as a triplet of qubits. This passage is particularly important because, continuing from \cite{Elze2020}, we will apply some perturbation on the system (for example, on its Hamiltonian) that triggers quantum behaviour.\\ 
Finally, we will illustrate how this classical system of three Ising spins that evolves exclusively through Ontological States behaves like a quantum system, evolving through superposition states, in particular. \\ This is an example of the potential implications of the CA Interpretation of Quantum Mechanics.
\end{changemargin}
\end{Abstract}
\tableofcontents

\bibliographystyle{ieeetr}

\chaptermark{Introduction}
\chapter*{Introduction}
It is well known, even amongst non expert people, that Quantum Mechanics (QM) describes a world that follows completely counter-intuitive rules, that has famous paradoxes as consequences, but that at the same time is the foundation of our description of the physical world that we can know through our senses.\\ Let's take for example the superposition principle: it applies well on wave functions but, since quantum objects behave both as wave and as particle, superposition must then apply on the latter too. So we have waves, that are also physical particles, which might be in two or more states at once. The peculiar thing is that, as strange and impossible as QM might seem, it provides scientists with extremely precise predictions. That is why it is now largely accepted that, at the order of magnitude of an atom, the laws of Physics are profoundly different from the ones of the macroscopic world and that the human mind struggles to fully understand.

There are of course many interpretations of QM, currently the most widely accepted  is the Copenhagen Interpretation: it states that, despite all the paradoxes, QM reflects how a quantum system actually behaves. Following the previous example, a particle can exist in a superposition of states, until it is observed. Once someone measures the particle, it collapses into a defined state because the act of measuring directly affects it.\\
Some other interpretations are instead trying to extend the deterministic properties from the classical world to the microscopic world, and one of these is the basis for this thesis. The so called Cellular Automata Interpretation of Quantum Mechanics by Gerard 't Hooft \cite{Hooft2014} claims that the Cellular Automata (CA) framework can be instrumental to the contextualization of Quantum Mechanics in a totally deterministic reality, without changing its formalism. According to this interpretation, particles evolve following the rules of CA, which we introduce later. Their states, by definition, are classical and thus deterministic and not in superposition: they are defined as Ontological States. The problem of measuring these states is solved by considering the observer and his/her instruments as a separate physical system which is equally deterministic, so the act of measuring is interpreted as the interaction between the two systems. Unfortunately, it is not yet known how to determine the Ontological States, in general. To overcome this obstacle, the Quantum Mechanics formalism works as a "\textit{template}" through which a scientist can work and get reliable results even without fully knowing what is happening at the microscopic level.

A Cellular Automaton is a discrete mathematical model that can be applied in various fields such as Mathematics, Biology, Theory of Computation, and Physics. \\ This model was first introduced in the 1940s by Stanislaw Ulam \cite{Ulam1952} and John von Neumann \cite{Neumann1951}: the former was working on the growth of crystals and the latter on the kinematic model, namely robots building each other. Both problems have in common being based on self-replicating systems, hence both authors decided to implement their own structure in a discrete framework.\\
The idea then took off and became popular later at the beginning of the 70s, when John Conway proposed the so called "\textit{Conway's Game of Life}" \cite{Gardener1970}\cite{Gardener1971}, a model of a population of "cells" with deterministic and discrete (both in space and in time) evolution rules: this model is considered a simple and primitive framework for the study of evolving systems. Another distinctive contribution to the spreading of possible applications was made by Stephen Wolfram, first publishing a study \cite{Wolfram1983} using CA properties to examine self-organization phenomena in statistical mechanics; here he gave a formal definition for CA and their properties (\cite{Wolfram1983} pag. 602)
\begin{quote}
Cellular automata are mathematical idealizations of
physical systems in which space and time are discrete,
and physical quantities take on a finite set of discrete
values. A cellular automaton consists of a regular uniform lattice (or "array"), usually infinite in extent, with a
discrete variable at each site ("cell"). The state of a cellular automaton is completely specified by the values of the
variables at each site. [\dots]
\end{quote}
Some decades later his book came out, "\textit{A new kind of science}" \cite{Wolfram2002}, where he extensively analyses how CA could be an alternative framework to address many issues from different scientific backgrounds.

We now know that Quantum Mechanics is one of these backgrounds. In 't Hooft's interpretation, the laws that are followed at the microscopic level are the discrete laws of Cellular Automata,  which exclude non classical phenomena such as superpositions. These are in fact interpreted as direct consequences of perturbations and limited precision in measurement.\\
This is not a brand new conception of how knowledge works, but in Philosophy the idea of the impossibility to know reality as it is for the human mind already existed. For example, in the XVIII century Kant in the "\textit{Critique of Pure Reason}" talked about reality as a \textit{Ding an sich}, \textit{thing-as-itself} \cite{Russell1945}\cite{Severino1996} that can not ever be known. In order to study the thing-as-itself at its best approximation, the human mind needs models and mind structures which can be shaped at its own pleasure, Kant calls this ensemble of models and mind structures as \textit{phenomenon}.\\
Following this parallelism, the real microscopic world, that 't Hooft calls Ontological, is what Kant calls \textit{thing-as-itself}, while Quantum Mechanics is the \textit{phenomenon} that we are researching, that 't Hooft instead calls \textit{template}, but its role remains the same.

Of course, it is not easy, if not currently impossible, to definitely prove that particles at microscopic level follow deterministic laws such as the ones followed by CA, or at least something similar to them. However, it is possible to prove that, within this framework, Quantum paradoxes come from measuring inaccuracies, and this is what is pursued in this thesis. \\
In the same spirit of \cite{Elze2020}, a system of three coupled Ising spins is considered. Such system, being classical, is interpreted as a Cellular Automaton, and its states are identified as Ontological States. Dynamics is then introduced by means of a classical operator that permutes the position of the spins; it has been proven that it is also possible to rewrite it using Pauli matrices. This means that this dynamics can be inserted in a Hilbert space and applied to a quantum system such as a chain of qubits. Once this is defined, we will try to prove that a QM feature such as superposition states, that do not exist on the Ontological level, arise if we make a perturbation on the system, both at the evolution operator level and at the Hamiltonian level.

The thesis is structured as follows: we will briefly describe the basic concepts of CA interpretation in Chapter 1 as exstensively explained in \cite{Hooft2014}, focusing on the notions that will be most needed later; in Chapter 2 there will be an extensive summary of the article \cite{Elze2020}, upon which this work is based: the article analyses a triplet of classical Ising spins and the permutation operators that can be applied on them, provides a quantum description of said operators, and finds a new finite Baker-Campbell-Hausdorff formula for the time evolution operator applied on the three spins chain.\\
In the 3rd and final Chapter we will continue where \cite{Elze2020} left, investigating how different kinds of perturbations applied either on the permutation operator or on the Hamiltonian can give to a classical system quantum features, such as superposition of states.
Finally, we draw some conclusions and open questions.

\addcontentsline{toc}{chapter}{Introduction}

\chaptermark{Chapter 1}
\chapter{Fundamentals of the Cellular Automata Interpretation of Quantum Mechanics}

\section{Definition of States}

In order to describe how states are defined in the CA interpretation, one must imagine reality divided into three levels: the superficial one, what we can actually see on a macroscopic order of magnitude, the deepest level which describes exactly the states of every single elementary particle or its underlying structure; finally, in the middle, there is our interpretation of the deepest level. These levels are defined as \cite{Hooft2014}\cite{Elze2020}:

\begin{figure}[h!]
\centering
\includegraphics[scale=0.25]{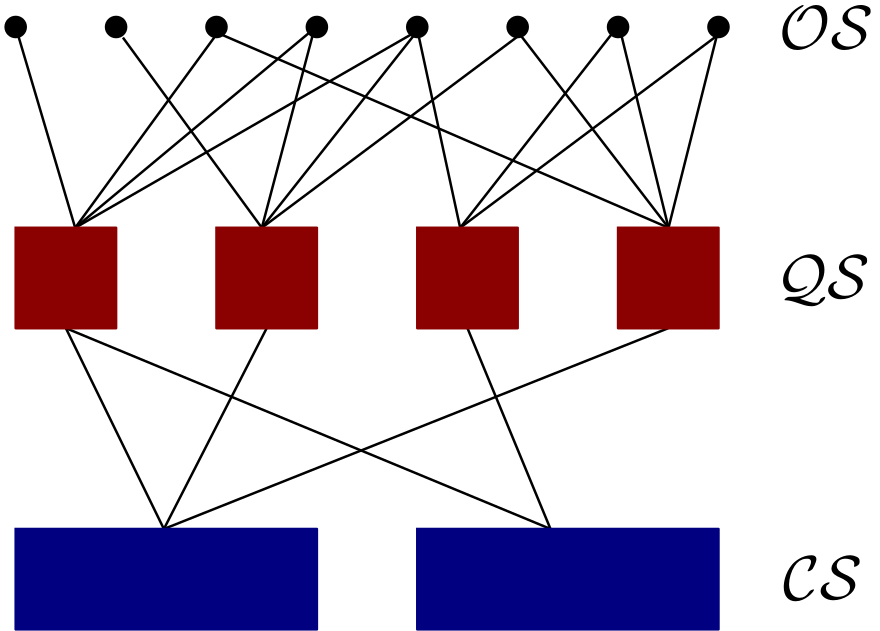}
\caption{The three levels of reality described in the CA interpretation. Notice how both $\mathcal{QS}$ and $\mathcal{CS}$ are superpositions of states from the level above, only in the $OS$ level there is no superposition. }\label{fig:3}
\end{figure}

\begin{itemize}

\item \textbf{Classical States}, $\mathcal{CS}$\\
As said before, these are the states we can see (not necessarily with our eyes); these are, of course, deterministic states, so to be more precise about the "macroscopic", this definition lasts as long as a closed system can be described without using the superposition formalism, so it concerns phenomena as they appear in measurements using macroscopic laboratory apparatus.

\item \textbf{Ontological States}, $\mathcal{OS}$\\
This is the deepest level. Since we want to postulate a deterministic behaviour, states that evolve in superposition states are forbidden at this level. What is being built is a whole deterministic reality, from all elementary particles to galaxies: the $\mathcal{OS}$ are the ones that shape the $\mathcal{CS}$ and since we do not have a detailed knowledge of all the $\mathcal{OS}$ of the Universe, Classical States are also defined as \textit{probabilistic distributions of the $\mathcal{OS}$ in the physical closed macrosystem}. This concept will be explained more clearly in the following Conservation Law.

\begin{figure}[h!]
\centering
\includegraphics[scale=0.3]{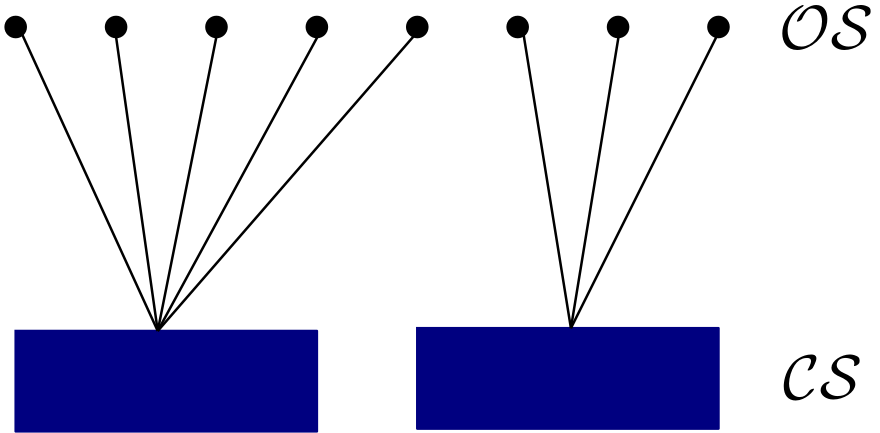}
\vspace{0.5cm}
\caption{Classical states as probabilistic distribution of the Ontological states, thus every line represents a probability.}\label{fig:stati}
\end{figure}

\begin{theorem*}[Conservation Law]
An ontological state evolves into an ontological state.
\end{theorem*}
This is not as trivial as it seems because it has two consequences: firstly it denies any possibility for an ontological state to evolve in a superposition state, so there is no chance superposition could be an actual physical state; secondly it says the whole reality is deterministic, and that means that classical state are ontological states too. To close the circle and connect ontological states to classical states we notice that in a QM experiment the probability of an outcome can be interpreted to be correlated to the probability to have some particular $\mathcal{OS}$ as initial condition \cite{Elze2020}.

\item \textbf{Quantum States}, $\mathcal{QS}$
\begin{equation}\label{qs}
|\psi\rangle=\sum_{A} \lambda_{A}|A\rangle, \quad \sum_{A}\left|\lambda_{A}\right|^{2} \equiv 1
\end{equation}

In \eqref{qs} the $\mathcal{QS}$ $|\psi\rangle$ is defined as a superposition of $\mathcal{OS}$ $|A\rangle$; the coefficients $\lambda_A$ of the linear combination need to respect the QM Born rule, so at this level all QM laws are followed as usual.
These aren't real states anymore, remember that superposition is not a reality, instead this is a pure mathematical trick that we need in order to get an idea what's happening on the $\mathcal{OS}$ scale. Gerard 't Hooft refers to them as \textit{templates}, and, to quote his exact words (\cite{Hooft2014} pag. 19):

\begin{quote}
A template is a quantum state of the form \eqref{qs} describing a situation where the probability to find our system to be in the ontological state $|A\rangle$ is $|\lambda^{2}|_A$.
\end{quote}

Thanks to this expedient, one can use all the QM formalism and its tools in particular the Born's rule and operators from the Hilbert space. 't Hooft cares to specify that these aren't the only available choices (for example another power law can be used instead of the Born's rule) and the only reason they are being maintained is the fact that they are the most convenient to use.
\end{itemize}

We mentioned the Hilbert space (HS): this is an important tool that works as a connection between the ontological states and the quantum formalism: $\mathcal{OS}$ do not individually belong to HS, but we may assume an orthonormal set of $\mathcal{OS}$ that can be a basis for an HS $\mathcal{\hat{H}}$: since $\mathcal{QS}$ are linear combinations of $\mathcal{OS}$, they therefore belong to $\mathcal{\hat{H}}$ as they should. In this basis we can build diagonal operators, whose eigenvalues define the physical properties of the state; these operators will be referred not as observables but as \textit{beables}, on the other hand, non-diagonal operators move the system from an ontological state to another and thus they are called \textit{changeables}: these will be discussed in the next section.\\
Talking about operators, since we are excluding superposition, consequently the collapse of a quantum state is not something that can happen anymore. A theory to explain what happens during measurements is to consider the act of measuring as a way to make two different systems interact. Although this explanation seems quite natural and elegant, it still presents a unresolved problem in the CA interpretation, the one concerning the interactions between different systems. In this thesis we will try to focus on this particular task. \\

\section{Dynamics}
\begin{figure}[h!]
\centering
\includegraphics[scale=0.4]{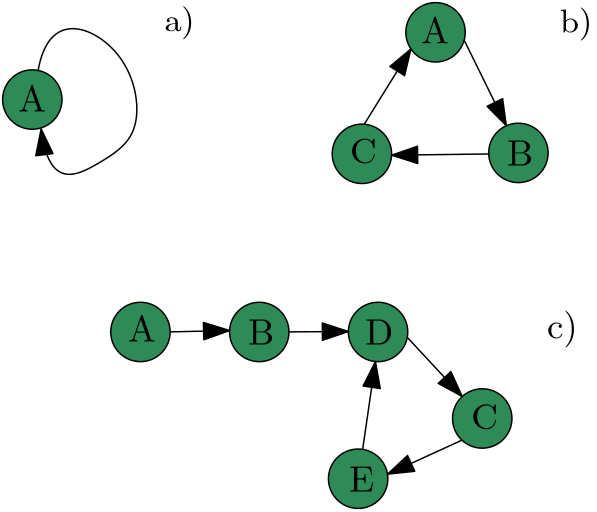}
\caption{a) One state system. b) Cogwheel model. c) System with information loss.}\label{fig:trans}
\end{figure}

In the previous section we defined \textit{changeables} as operators that describe how a system can change its state. These operators act as \textit{permutations} on a finite set of $N$ states, $\{ |A\rangle, |B\rangle, |C\rangle\dots \} $ so that there can be three different scenarios:
\begin{enumerate}
\item A system doesn't change its state, so its $\mathcal{OS}$ are mapped onto themselves \ref{fig:trans}a
\begin{equation}
|A\rangle\rightarrow|A\rangle, \qquad |B\rangle\rightarrow|B\rangle \dots
\end{equation}
\item The elements of a set are mapped onto the same set. This is referred as \textit{Cogwheel Model} (in fig.\ref{fig:trans}b it is clear why):
\begin{equation}
|C\rangle\rightarrow|A\rangle\rightarrow|B\rangle\dots
\end{equation}
A set of N elements allows N! permutations; however, here each state happens only once in a cyclic update of the system, which evolves by following exactly one precisely defined permutation rule (we do not  consider a changing set of states at present). 
\item Two or more $\mathcal{OS}$ of a system evolve into the same $\mathcal{OS}$: 
\begin{equation}
|A\rangle\rightarrow|B\rangle\rightarrow|D\rangle\rightarrow|C\rangle\rightarrow|E\rangle\rightarrow|D\rangle
\end{equation}
This is a rather complex event, referred as \textit{information loss}, since the first set of states that do not evolve into a closed cycle (fig.\ref{fig:trans}c) are realized once and then lost. This case has quite interesting developments and applications but these are not of our interest here. More informations are in \cite{Hooft2014} chap. 7.
\end{enumerate}

A peculiarity of this model is that time is no longer continuous but it is represented by discrete and integer numbers. How it is possible to relate  the discrete CA time to the continuous time in QM is explained in Appendix A. Since the description of time changes, of course its meaning cannot remain the same: in fact, in this formalism time is not to be considered something that flows continuously and regardless of any physical event. Instead, time is now strictly connected to state change and both events happen simultaneously, the beat of the clock and the state transition. To quote from \cite{Elze2002} and \cite{Elze2003}:
\begin{quote}
Our construction of a reparametrization invariant "time" is motivated by the observation that "time passes" when there is an observable change, which is localized by the observer.
\end{quote}

Let's have a closer look to a simple three-states system to understand better how permutations work.\\

\section{The Cogwheel model}\label{cogwheel}
\begin{figure}[h!]
\centering
\includegraphics[scale=0.5]{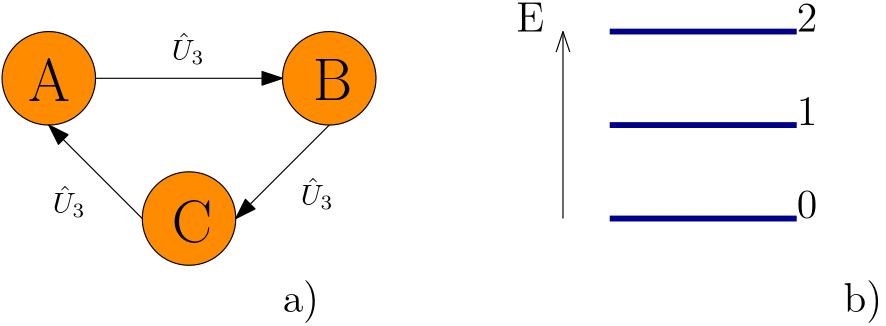}
\vspace{0.5cm}
\caption{a) A three-state cogwheel model and b) its energy levels with respective eigenstates.}\label{fig:cog}
\end{figure}

Let $|A\rangle, |B\rangle$, and $ |C\rangle$ be the states of a deterministic system \cite{Hooft2014}, its time evolution is (see fig \ref{fig:cog}):
\begin{equation}
|A\rangle \rightarrow |B\rangle \rightarrow |C\rangle \rightarrow |A\rangle \dots
\end{equation}

For every transition the clock ticks and time increases by an increment $T$, let us represent this event by a permutation operator $\hat{U}_{3}$:
\begin{align}
\hat{U}_{3}\,|A\rangle=|B\rangle,\nonumber\\
\hat{U}_{3}\,|B\rangle=|C\rangle,\nonumber\\
\hat{U}_{3}\,|C\rangle=|A\rangle.
\end{align}

It is possible to write $\hat{U}_{3}$ in matrix form: the most convenient way is to build a 3x3 matrix written in the auxiliary $\mathbb{R}^3$ basis, so that:
\begin{align}
|A\rangle=\begin{pmatrix}
1&0&0
\end{pmatrix}^t,\quad
&|B\rangle=\begin{pmatrix}
0&1&0
\end{pmatrix}^t,\quad
|C\rangle=\begin{pmatrix}
0&0&1
\end{pmatrix}^t, \nonumber\\
&\hat{U}_{3}=\begin{pmatrix}
0 & 0 & 1 \\
1 & 0 & 0 \\
0 & 1 & 0
\end{pmatrix}.\
\end{align}
This matrix has a couple of distinctive properties, typical of permutations:
\begin{equation}\label{prop}
\hat{U}_{N}^{\dagger}\hat{U}_{N}=\mathbf{Id},\qquad
\hat{U}_{N}^{N}=\mathbf{Id};
\end{equation}
in particular the second property comes from the fact that cogwheel models usually represent cyclic transformations meaning that, as the fig \ref{fig:cog} shows, after $N$ ticks of the clock the system returns to its original state, in this case we have $N=3$.\\
Now the next useful step would be to find an Hamiltonian for this system. In order to do this it is necessary to diagonalize the $\hat{U}_{3}$ matrix:
\begin{equation}
\hat{U}_{diag}=\begin{pmatrix}
1 & 0 & 0\\
0 & e^{i \frac{-2\pi}{3}} & 0\\
0 & 0 & e^{i \frac{-4\pi}{3}}\end{pmatrix},
\end{equation}
where the eigenvalues are three complex roots of 1 and the normalized eigenvectors are:
\begin{equation}
|0\rangle=\frac{1}{\sqrt{3}}\begin{pmatrix}
1\\
1\\
1
\end{pmatrix},\quad
|1\rangle=\frac{1}{\sqrt{3}}\begin{pmatrix}
1\\
e^{\frac{-i2\pi}{3}}\\
e^{\frac{i2\pi}{3}}
\end{pmatrix},\quad
|2\rangle=\frac{1}{\sqrt{3}}\begin{pmatrix}
1\\
e^{\frac{i2\pi}{3}}\\
e^{\frac{-i2\pi}{3}}
\end{pmatrix},
\end{equation}
so that:
\begin{equation}
\hat{U}_3|0\rangle=|0\rangle, \quad
\hat{U}_3|1\rangle=e^{\frac{-i2\pi}{3}}|1\rangle, \quad
\hat{U}_3|2\rangle=e^{\frac{-i4\pi}{3}}|2\rangle.
\end{equation}
Now it comes naturally to write the operator also as $\hat{U}_{3}=e^{-i \hat{H} T}\,\,$, with:
\begin{equation}\label{diag}
\hat{H}=\frac{2\pi}{3T}\begin{pmatrix}
0&0&0\\
0&1&0\\
0&0&2\\
\end{pmatrix};
\end{equation}
in this basis where the Hamiltonian is diagonal. The energy level are evidently as shown (fig.\ref{fig:cog}b). We can now go back to the original ontological basis $\{|A\rangle, |B\rangle, |C\rangle\}$, performing a base change with matrix $D$:
\begin{equation}
\hat{H}_{aux}=D^{\dagger}\hat{H}D, \qquad\ D=\frac{1}{\sqrt{3}}\left(\begin{array}{ccc}
1 & 1 & 1 \\
1 & e^{i \frac{2 \pi}{3}} & e^{-i \frac{12 \pi}{3}} \\
1 & e^{-i \frac{2 \pi}{3}} & e^{i \frac{2 \pi}{3}}
\end{array}\right);
\end{equation}
\begin{equation}\label{aux}
\hat{H}_{aux}=\frac{2 \pi}{3 T}\begin{pmatrix}
1 & \kappa & \kappa^{*} \\
\kappa^{*} & 1 & \kappa \\
\kappa & \kappa^{*} & 1
\end{pmatrix}, \quad  \kappa=-\frac{1}{2}+\frac{i \sqrt{3}}{6}.
\end{equation}
What has been done is to write an operator that acts on ontological states in the form of a time evolution operator in Quantum Mechanics, $\hat{U}_{3}(t)=e^{-i \hat{H} t}\,\,$, with the only difference that $t$ in this case must be an integer multiple of $T$. It is thus possible to build a \textit{template} superposition state $\psi(t)=a(t)|A\rangle+b(t)|B\rangle+c(t)|C\rangle$ that satisfies Schr\"odinger equation:
\begin{equation}
\frac{\mathrm{d}}{\mathrm{d} t}|\psi\rangle=-i \hat{H}|\psi\rangle.
\end{equation}
This procedure works for a general cogwheel model with $N$ states; here we have an example of how one can use mathematical tools typical of Quantum Mechanics for a different, deterministic and discrete model. An actual physical system with a limited number of states that can be assimilated to a cogwheel model are the electrons in the external shell of an atom: under the effect of an electromagnetic field, the Stark and the Zeeman effect separate all the energy levels with the result that the atom has a limited number of distinct spectroscopic configurations available and very specific paths to transition between them: these assumptions seems to be in accordance with the premises of our ontological model.
The only apparent problem is caused by the discrete time scale. However it is just a false difference, because, although there is a restriction on the value of $t$, since it must be an integer multiple of $T$. However $T$ itself doesn't have to be an integer, but it can be as small as it is needed by the scale of the system. For example, it could be useful to set as a time scale (see Appendix A) $l=10^{-43}\,\,s$, assuming that the Planck time is the smallest meaningful unit of time. Most physical events happen in much larger time scale, so that such a small time step is suitable for an approximation to the continuum: this allows to have negligible differences between an ontological model description and a quantum mechanical one.\\
\\
To end this Chapter, it needs to be remarked that until now we have considered only isolated, non interacting problems: the theory concerning interacting systems is still a work in progress. In the next Chapters we will try to introduce this possibility applying the permutation dynamics to a chain of three-coupled Ising spins.

\chaptermark{Chapter 2}
\chapter{Qubit exchange interactions from permutations of classical bits}
\section{Permutations of Classical Bits with a QM Hamiltonian}

As anticipated earlier, let's summarize the work done in \cite{Elze2020}, concerning a classical description of a triplet of Ising spins.\\
Ising spins are quite suitable for this purpose thanks to their two-state function that can be described classically; they can in fact represent generic Boolean variables, binary bits; also this behaviour is reminiscent of quantum spins and this will be relevant later. \\
\\
Given a chain of three coupled Ising spins $\{1,\, 2, \, 3\}$, a simple transformation for this system would be a sign exchange between two spins, and that's something easily represented by a permutation operator $\hat{P}$:
\begin{equation}
\hat{P}_{ij}|s_i,\,s_j\rangle=|s_j,\,s_i\rangle;
\end{equation}
being $|s_i,\,s_j\rangle$ the ontological state of two spins of the system, where $i,j=1,2,3$ and the value $s_i$ can be equal to $\uparrow$ or $\downarrow$, so that, for example, $\hat{P}_{12}$ swaps the spin $1$ and the spin $2$ of the system $|s_1,\,s_2\rangle$:  
\begin{equation}
\hat{P}_{12}|\uparrow\,\downarrow\,\rangle=|\downarrow\,\uparrow\,\rangle.
\end{equation}
We still need to clarify matters of numerical representation.\\
\\
As anticipated earlier, spins are usually considered as quantum objects, it comes in handy to use quantum representation: given the set of Pauli matrices,
\begin{equation}
\bm{\sigma}=
\begin{pmatrix}
\sigma_x & \sigma_y & \sigma_z
\end{pmatrix}^t=
\begin{pmatrix}
\begin{pmatrix}
0 & 1 \\
1 & 0
\end{pmatrix} &
\begin{pmatrix}
0 & -i \\
i & 0
\end{pmatrix} &
\begin{pmatrix}
1 & 0 \\
0 & -1
\end{pmatrix}
\end{pmatrix}^t,
\end{equation}
it is possible to use eigenvalues and eigenstates of the matrix $\sigma_z$ in relation to the states of an Ising spin:
\begin{itemize}
\item[-]state: $|\uparrow\,\rangle$, eigenvalue: $+1$, eigenvector: $\begin{pmatrix}
1\\0
\end{pmatrix}$
\item[-]state: $|\downarrow\,\rangle$, eigenvalue: $-1$, eigenvector: $\begin{pmatrix}
0\\1
\end{pmatrix}$.
\end{itemize}
Now the permutation operator $\hat{P}_{ij}$ can be expressed as a QM tensor:
\begin{equation}\label{qm}
\hat{P}_{ij}=\frac{1}{2}\left(\bm{\sigma}_{i} \cdot \bm{\sigma}_{j}+\mathbf{Id}\right).
\end{equation}
Let's see how this works using a test state $|s_1,\,s_2\rangle=|\uparrow\,\downarrow\,\rangle$:
\begin{align}
\hat{P}_{12}|s_1,\,s_2\rangle&=\frac{1}{2}\left(\bm{\sigma}_{1} \cdot \bm{\sigma}_{2}+\mathbf{Id}\right)\cdot|\uparrow\,\downarrow\,\rangle=\frac{1}{2}\left(\begin{pmatrix}
\sigma_x\\ \sigma_y \\ \sigma_z
\end{pmatrix}_{1} \cdot \begin{pmatrix}
\sigma_x\\ \sigma_y \\ \sigma_z
\end{pmatrix}_{2}+\mathbf{Id}\right)\cdot|\uparrow\,\downarrow\,\rangle \nonumber \\
&=\frac{1}{2}\left(\sigma_{x1}\sigma_{x2}+\sigma_{y1}\sigma_{y2}+\sigma_{z1}\sigma_{z2}+\mathbf{Id}\right)\cdot|\uparrow\,\downarrow\,\rangle \nonumber \\
&=\frac{1}{2}\left(\sigma_{x1}|\uparrow\,\rangle\sigma_{x2}|\downarrow\,\rangle+\sigma_{y1}|\uparrow\,\rangle\sigma_{y2}|\downarrow\,\rangle+\sigma_{z1}|\uparrow\,\rangle\sigma_{z2}|\downarrow\,\rangle+|\uparrow\,\downarrow\,\rangle \right) \nonumber \\
&=\frac{1}{2}\left(|\downarrow\,\rangle|\uparrow\,\rangle+i|\downarrow\,\rangle(-i|\downarrow\,\rangle)+|\uparrow\,\rangle(-|\downarrow\,\rangle)+|\uparrow\,\downarrow\,\rangle \right) \nonumber \\
&=|\downarrow\,\uparrow\,\rangle.
\end{align} 
We remark that writing $\hat{P}_{ij}$ as such a tensor allows us to insert the Ising spins triplets in a Quantum Mechanics context.\\
\\
It will be useful to express $\hat{P}_{ij}$ in  matrix form too, with a $\mathbb{R}^4$ orthonormal basis:
\begin{equation*}
|\uparrow\,\uparrow\,\rangle=\begin{pmatrix}
1\\0\\0\\0
\end{pmatrix}=|1\rangle,
\quad|\uparrow\,\downarrow\,\rangle=\begin{pmatrix}
0\\1\\0\\0
\end{pmatrix}=|2\rangle,
\quad|\downarrow\,\uparrow\,\rangle=\begin{pmatrix}
0\\0\\1\\0
\end{pmatrix}=|3\rangle,
\quad|\downarrow\,\downarrow\,\rangle=\begin{pmatrix}
0\\0\\0\\1
\end{pmatrix}=|4\rangle
\end{equation*}
so that each element $(k,l)$ of the matrix will be the scalar product $\hat{P}_{ij}|k\rangle\cdot |l\rangle$.\\ Let's apply this to $\hat{P}_{12}$:
\begin{equation}\label{matrix}
\hat{P}_{12}=\begin{pmatrix}
1 & 0 & 0 & 0 \\
0 & 0 & 1 & 0 \\
0 & 1 & 0 & 0 \\
0 & 0 & 0 & 1
\end{pmatrix}.
\end{equation}
This equation allows us to verify that $\hat{P}_{ij}$ satisfies the properties seen in eq \eqref{prop} which are ($N=2$ in this case):
\begin{equation}\label{properties}
\hat{P}_{ij}^{\dagger}\hat{P}_{ij}=\mathbf{Id},\qquad
\hat{P}_{ij}^{2}=\mathbf{Id}.
\end{equation}
Also eq. \eqref{matrix} is useful to demonstrate commutation properties of the operator $\hat{P}_{ij}$: 
\begin{equation}
\left[\hat{P}_{ij},\hat{P}_{jk}\right]\neq0, \qquad i\neq k
\end{equation}

\section{Dynamics for a triplet of coupled Ising spins}\label{sez}

In order to extend the $\hat{P}_{ij}$ operator to a triplet of bits $|s_1\,s_2\,s_3\rangle$ it is enough to concatenate two different ones $\hat{P}_{ij}\hat{P}_{kl}$ , in particular we will be using $\hat{P}_{12}\hat{P}_{23}$:
\begin{align}
\hat{P}_{12}\hat{P}_{23}|s_1\,s_2\,s_3\rangle=|s_3\,s_1\,s_2\rangle \rightarrow 
\hat{P}_{12}\hat{P}_{23}|s_3\,s_1\,s_2\rangle=|s_2\,s_3\,s_1\rangle \rightarrow \nonumber \\ \rightarrow
\hat{P}_{12}\hat{P}_{23}|s_2\,s_3\,s_1\rangle=|s_1\,s_2\,s_3\rangle. 
\end{align}\\
The next steps will follow Sec. \ref{cogwheel} proceeding. We aim to write at the end an equation for the Hamiltonian $\hat{H}$ of this system.\\
\\ 
This system is in fact something slightly more complex than the cogwheel model of section \ref{cogwheel}. We still have a cyclic permutation that every $N=3$ discrete time steps $T$ takes the system back to its initial state, as shown before. In this case too it is possible to write the permutation operator as a Quantum Mechanics time evolution operator:
\begin{equation}\label{expon}
\hat{P}_{12}\hat{P}_{23}\stackrel{\text{def}}{=}\hat{U}=e^{-i\hat{H}T}.
\end{equation}
The matrix form for $\hat{U}$ will be written in the $\mathbb{R}^8$ auxiliary basis $\{|1\rangle\dots|8|\rangle\}$ , since we have $2^3$ possible ontological states that are ordered in the following way:
\begin{equation}\label{basis}
|\uparrow\uparrow\uparrow\,\rangle,\quad
|\uparrow\uparrow\downarrow\,\rangle,\quad
|\uparrow\downarrow\uparrow\,\rangle,\quad
|\downarrow\uparrow\uparrow\,\rangle,\quad
|\downarrow\downarrow\uparrow\,\rangle,\quad
|\downarrow\uparrow\downarrow\,\rangle,\quad
|\uparrow\downarrow\downarrow\,\rangle,\quad
|\downarrow\downarrow\downarrow\,\rangle.
\end{equation}
Then, following the same logic of \eqref{matrix}, the matrix will come out like this:
\begin{equation}\label{bigm}
\hat{U}=\begin{pmatrix}
1&0&0&0&0&0&0&0\\
0&0&1&0&0&0&0&0\\
0&0&0&1&0&0&0&0\\
0&1&0&0&0&0&0&0\\
0&0&0&0&0&1&0&0\\
0&0&0&0&0&0&1&0\\
0&0&0&0&1&0&0&0\\
0&0&0&0&0&0&0&1
\end{pmatrix}.
\end{equation}
This matrix is actually quite big to work with, but there is a peculiarity that can be used to simplify it: the operator $\hat{U}$ keeps the number of up- and down-spins in the system conserved, and as a consequence of this, the matrix \eqref{bigm} is made of 4 different blocks, each for every distinct combination of three up and down-spins. The first block is the element $(1,1)$ that represents the state with three up-spins, the second block is a 3x3 submatrix that represents the three states with two up-spins and one down-spin, the third one is the same as the second one but for the three states with two down-spins and one up-spin and finally the last one is the element $(8,8)$ linked to the remaining state with three down-spins:
\begin{equation}
\hat{U}=\begin{pmatrix}
1 & & &\\
& U_{3} &  &\\
&  & U_{3} & \\
& & & 1
\end{pmatrix},\qquad
\hat{U}_3=\begin{pmatrix}
0&1&0\\
0&0&1\\
1&0&0
\end{pmatrix}.
\end{equation}
This diagonal block form is convenient to diagonalize the $\hat{U}$ matrix: eigenvalues and eigenvectors for the two 1x1 blocks are trivial, while for a single $\hat{U}_3$ they are (vectors are already normalized):
\begin{equation*}
\lambda_1=1,\qquad 
\lambda_2=e^{\frac{-i2\pi}{3}},\qquad 
\lambda_3=e^{\frac{-i4\pi}{3}}
\end{equation*}
\begin{equation}\label{eigen}
\mathbf{v}_1=\frac{1}{\sqrt{3}}\begin{pmatrix}
1\\
1\\
1
\end{pmatrix},\quad
\mathbf{v}_2=\frac{1}{\sqrt{3}}\begin{pmatrix}
1\\
e^{\frac{-i2\pi}{3}}\\
e^{\frac{i2\pi}{3}}
\end{pmatrix},\quad
\mathbf{v}_3=\frac{1}{\sqrt{3}}\begin{pmatrix}
1\\
e^{\frac{i2\pi}{3}}\\
e^{\frac{-i2\pi}{3}}
\end{pmatrix}
\end{equation}
It thus becomes immediate to write $\hat{H}$ from eq. \eqref{expon} in a diagonal form, with the eigenvectors of $\hat{U}$ chosen as basis Knowing that also for $\hat{U}_3$ it is true that $\hat{U}_3^3=\mathbf{Id}$, the diagonal representation for $\hat{H}$ is, remembering eq. \eqref{diag}:
\begin{equation}\label{ham}
\hat{H}=\frac{2\pi}{3T}\text{diag}\left\{0,0,1,2,0,1,2,0\right\}
\end{equation}

\subsection{$\hat{H}$ as a function of permutation operators}\label{quelladopo}
The goal here is to write the Hamiltonian $\hat{H}$ as an operator to apply to the vectors of the auxiliary basis \eqref{basis}. First we need a basis change on the matrix \eqref{ham}, here too it is convenient to work on the single block:
\begin{equation}
\hat{H}_3=D^{\dagger}\hat{U}_3D,
\end{equation}
where $D$ is the basis change matrix built with the set of eigenvectors \eqref{eigen}, and the result that comes out is, as expected, equal to eq. \eqref{aux}:
\begin{equation}\label{aux2}
\hat{H}_3=\frac{2 \pi}{3 T}\begin{pmatrix}
1 & \kappa & \kappa^{*} \\
\kappa^{*} & 1 & \kappa \\
\kappa & \kappa^{*} & 1
\end{pmatrix}, \quad  \kappa=-\frac{1}{2}+\frac{i \sqrt{3}}{6}.
\end{equation}
Now that we have an Hamiltonian matrix written in the auxiliary basis, let's see how it acts on the ontological states. Since we are still working on blocks, it is useful to rewrite \eqref{basis} as:
\begin{equation}\label{up}
|\uparrow \uparrow \downarrow \,\rangle=\begin{pmatrix}
1\\0\\0
\end{pmatrix}=|\alpha\rangle,
\quad|\uparrow \downarrow \uparrow\,\rangle=\begin{pmatrix}
0\\1\\0
\end{pmatrix}=|\beta\rangle,
\quad|\downarrow\uparrow \uparrow\,\rangle=\begin{pmatrix}
0\\0\\1
\end{pmatrix}=|\gamma\rangle,
\end{equation}
for the first 3x3 block and similarly for the second:
\begin{equation}\label{down}
|\downarrow \downarrow \uparrow \,\rangle=\begin{pmatrix}
1\\0\\0
\end{pmatrix}=|\alpha\rangle,
\quad|\downarrow \uparrow \downarrow\,\rangle=\begin{pmatrix}
0\\1\\0
\end{pmatrix}=|\beta\rangle,
\quad|\uparrow\downarrow \downarrow\,\rangle=\begin{pmatrix}
0\\0\\1
\end{pmatrix}=|\gamma\rangle.
\end{equation}
The $\mathcal{OS}$ from both sets have been labelled with the same kets, as the matrix $\hat{H}_{3}$ applied on $\{|\alpha\rangle,|\beta\rangle,|\gamma\rangle\} $ gives the same result regardless whether these vectors come from \eqref{up} or \eqref{down}.\\
The last step that needs to be done, before applying $\hat{H}_{3}$ on the $\{|\alpha\rangle,|\beta\rangle,|\gamma\rangle\} $ set, is to write down the following relations:
\begin{equation}
|\alpha\rangle=\hat{P}_{23}|\beta\rangle=\hat{P}_{13}|\gamma\rangle, \qquad
|\beta\rangle=\hat{P}_{12}|\gamma\rangle= \hat{P}_{23}|\alpha\rangle, \qquad
|\gamma\rangle=\hat{P}_{12}|\beta\rangle=\hat{P}_{13}|\alpha\rangle,
\end{equation}
following from \eqref{up} and \eqref{down} by simply performing the indicated permutations of spin variables. Thus, we obtain:
\begin{equation}
\begin{aligned}
\hat{H}_3|\alpha\rangle &=\frac{2 \pi}{3 T}\left(|\alpha\rangle+\kappa^*|\beta\rangle+\kappa|\gamma\rangle\right)
\equiv \frac{2 \pi}{3 T}\left(\mathbf{Id}+\kappa^*\hat{P}_{23}+\kappa \hat{P}_{12}\right)|\uparrow \uparrow \downarrow\,\rangle, \\
\hat{H}_3|\beta\rangle &=\frac{2 \pi}{3 T}\left(\kappa^*|\alpha\rangle+|\beta\rangle+\kappa|\gamma\rangle\right)
\equiv \frac{2 \pi}{3 T}\left(\kappa^* \hat{P}_{23}+\mathbf{Id}+\kappa \hat{P}_{12}\right)|\uparrow \downarrow \uparrow\,\rangle,\\
\hat{H}_3|\gamma\rangle &=\frac{2 \pi}{3 T}\left(\kappa|\alpha\rangle+\kappa^*|\beta\rangle+|\gamma\rangle\right)
\equiv \frac{2 \pi}{3 T}\left(\kappa \hat{P}_{31}+\kappa^{*} \hat{P}_{12}+\mathbf{Id}\right)|\downarrow \uparrow \uparrow\,\rangle.
\end{aligned}
\end{equation}
We are now closer to obtain an Hamiltonian operator to be applied on all 8 ontological states from \eqref{basis}. In order to reach this result, it is useful to observe that the rows in matrix \eqref{aux2} are are related to each other by cyclic permutations of their elements; using the notation $\{x, y, z\}$ for the three rows we have:
\begin{equation}\label{rotations}
\hat{P}_{13} \hat{P}_{23}|x, y, z\rangle=|y, z, x\rangle,\qquad\left(\hat{P}_{13} \hat{P}_{23}\right)^{2}|x, y, z\rangle=|z, x, y\rangle,
\end{equation}
that are respectively anticlockwise and clockwise permutations. This allows us to write \eqref{ham} in an explicit form as:
\begin{equation}\label{result}
\hat{H}=\frac{2 \pi}{3 T}\left(\mathbf{Id}+\kappa^{*} \hat{P}_{13} \hat{P}_{23}+\kappa\left(\hat{P}_{13} \hat{P}_{23}\right)^{2}\right),
\end{equation}
or, since $\left(\hat{P}_{13}\hat{P}_{23}\right)^2=\hat{P}_{23}\hat{P}_{13}$,
\begin{equation}\label{result2}
\hat{H}=\frac{2 \pi}{3 T}\left(\mathbf{Id}+\kappa^{*} \hat{P}_{13} \hat{P}_{23}+\kappa \hat{P}_{23} \hat{P}_{13}\right),
\end{equation}
which presents a main result in \cite{Elze2020}. To end this section, let's point out some properties of eq. \eqref{result}:
\begin{itemize}
\item[-]As an Hamiltoniam, $\hat{H}$ is of course self-adjoint;
\item[-] $\left(\hat{P}_{13}\hat{P}_{23}\right)^3=\mathbf{Id}$, as it should be;
\item[-] $\left(\hat{P}_{13}\hat{P}_{23}\right)^{-1}=\left(\hat{P}_{13}\hat{P}_{23}\right)^{\dagger}=\left(\hat{P}_{13}\hat{P}_{23}\right)^2$;
\item[-] the last thing worth noting is that all operators composing \eqref{result} commute with each other, in particular $\left[\hat{P}_{23} \hat{P}_{13}, \hat{P}_{13} \hat{P}_{23}\right]=\mathbf{Id}-\mathbf{Id}$, a consequence of the precedent point.
\end{itemize}

The next section will be about the consequences of equations \eqref{result} or \eqref{result2}.

\section{A BCH formula with a qubit Hamiltonian for bits}
Our previous results provide a solution for a formal problem that can now be unravelled. \\
The permutation operator defined in the beginning of this Chapter allows us to write:
\begin{equation}\label{exp}
\hat{P}_{ij}=i\exp\left[\frac{-i\hat{P}_{ij}\pi}{2}\right].
\end{equation}
However this is not easily generalized for the three-spins case where the operator is a combination of two permutations: we used $\hat{P}_{12}\hat{P}_{23}$ to which \eqref{exp} cannot be applied, because, as previously verified, $\hat{P}_{12}$ and $\hat{P}_{23}$ do not commute with each other.\\
So the problem would be to find some $Z$ so that:
\begin{equation}\label{problem}
\exp\left[\hat{P}_{12}\right]\exp\left[\hat{P}_{23}\right]=\exp\left[Z\right].
\end{equation}
It does exist a specific solution for this kind of problem, which is the Baker-Campbell-Hausdorff (BCH) formula, here written to the 4th order in increasingly complex commutators:
\begin{equation}
Z=X+Y+\frac{1}{2}[X, Y]+\frac{1}{12}([X,[X, Y]]+[Y,[Y, X]])-\frac{1}{24}[Y,[X,[X, Y]]]+\ldots ;
\end{equation}
but, while every term is known, it is difficult to verify convergence or prove divergence of the series. There are some exceptional cases and \eqref{problem} can be considered one of them. In fact, as a consequence of section \ref{sez}, a new BCH formula for $\hat{P}_{12}\hat{P}_{23}$ is:
\begin{equation}\label{nuovo}
i^{2} \exp \left(-i \frac{\pi}{2} \hat{P}_{12}\right) \exp \left(-i \frac{\pi}{2} \hat{P}_{23}\right)=\exp \left(-i \frac{2 \pi}{3}\left(\mathbf{Id}+\kappa \hat{P}_{23} \hat{P}_{13}+\kappa^{*} \hat{P}_{13} \hat{P}_{23}\right)\right).
\end{equation}

Another particular case is the permutation operator $\hat{P}_{23}\hat{P}_{12}\hat{P}_{34}$ applied to a system of four Ising spins: in \cite{Elze20202} a similar procedure was followed and another BCH formula has been obtained.
\\
\\
What has been achieved here is to describe a classical, deterministic system with an Hamiltonian that, thanks to eq. \eqref{qm}, can be inserted in the context of an Hilbert space:
\begin{equation}\label{quantistica}
\hat{P}_{12}\hat{P}_{23}=\frac{1}{4}\left(\bm{\sigma}_{1} \cdot \bm{\sigma}_{2}+\bm{\sigma}_1\cdot\bm{\sigma}_3+\bm{\sigma}_2\cdot\bm{\sigma}_3+\mathbf{Id}\right).
\end{equation}
In this case we have quantum objects like qubits, that have the same double state characteristic of Ising spins, and that make them eligible systems for the operator
\eqref{nuovo} to be applied on: in this way, \eqref{nuovo}  appears as a legitimate \textit{quantum~mechanical~operator}.\\ \\
So a quantum system of qubits can evolve through deterministic $\mathcal{OS}$, in particular, avoiding superposition states: this is actually true as long as every coefficient of \eqref{nuovo} is precisely determined without any pertubation. Let's see what happens if just one permutation operator is perturbed:
\begin{equation}\label{pert}
i \exp \left(-i \frac{\pi}{2}(1+\varepsilon) \hat{P}_{i j}\right)=\hat{P}_{i j}-i \frac{\pi}{2} \varepsilon \cdot \mathbf{Id}+\mathrm{O}\left(\varepsilon^{2}\right), \quad 0<\epsilon \ll 1.
\end{equation}
This is an operator that generally will produce superposition states, which would be incompatible with $\mathcal{OS}$ evolution. \\
\\
To quote \cite{Elze2020}:
\begin{quote}
We conclude that an only \textit{approximately known ontological Hamiltonian must lead to misinterpretation}, namely that the system under study behaves \textit{quantum mechanically}, due to the presence of superpositions of $\mathcal{OS}$.
\end{quote}

The next Chapter will be focused on this particular phenomenon.We will also provide a detailed explanation related to eq.\eqref{pert}.

\chapter{Consequences of perturbations on a classical chain of Ising spins}
In this Chapter we want to elaborate the matter of perturbations, the consequences of which were mentioned at the end of Chapter 2.\\
\\
At the beginning of this thesis the fact has been stated that, since ontological states obey to deterministic mechanical laws, an $\mathcal{OS}$ built as a superposition of other $\mathcal{OS}$ cannot exist. But it is possible to use Quantum Mechanics as a template, so in the context of quantum states one eventually will use superposition states. \\Thus the question is: when do superposition states come out and what do they represent?\\
\\
One example can be the case seen illustrated by eq.\eqref{pert}. An approximately determined Hamiltonian can generate a state that does not belong anymore to an ontological framework.\\
We will explain the implications related to eq.\eqref{pert} in the context of our previously analysed system.

\section{Perturbation of permutation operator $\hat{P}_{12}\hat{P}_{23}$}\label{diagonal}
In order to see superposition states arise, we will work with the previous system of three Ising spins $|s_1\,s_2\,s_3\rangle$ and with the evolution operator $\hat{P}_{12}\hat{P}_{23}$. Let's remind that it can be written:
\begin{equation}\label{primo}
\hat{P}_{12}\hat{P}_{23}=
-\exp \left(-i \frac{\pi}{2} \hat{P}_{12}\right)\exp \left(-i \frac{\pi}{2} \hat{P}_{23}\right),
\end{equation}
thanks to the \eqref{properties} properties. Also:
\begin{equation}\label{secondo}
\hat{P}_{12}\hat{P}_{23}=\exp \left(-i\hat{H}T\right)=
\exp{\left(-i\frac{2\pi}{3}\left(\mathbf{Id}+\kappa \hat{P}_{23} \hat{P}_{13}+\kappa^{*} \hat{P}_{13} \hat{P}_{23}\right)\right)},
\end{equation} 
thanks to the BCH formula \eqref{nuovo}.\\
Let's follow the example of equation \eqref{pert} and apply a first perturbation to the system: we multiply the exponent by a factor $(1+\varepsilon)$; from here through all the Chapter we will set the parameter $\varepsilon$ to be real and $0<\varepsilon\ll1$.\\ Now, this can be done in two ways and they both provide two different results: we can perturb the exponents in \eqref{primo} or the exponent in \eqref{secondo}.\\
\\
Starting with \eqref{primo}, the perturbed operator comes out like this:
\begin{equation}
-\exp\left(-i\frac{\pi}{2}\hat{P}_{12}\left(1+\varepsilon\right)\right)\exp\left(-i\frac{\pi}{2}\hat{P}_{23}\left(1+\varepsilon\right)\right);
\end{equation}
by using eq.\eqref{primo} the exponent in $\varepsilon$ can be isolated so that a Taylor expansion in $\varepsilon$ can be applied:
\begin{align}\label{first}
&-\exp\left(-i\frac{\pi}{2}\hat{P}_{12}\left(1+\varepsilon\right)\right)\exp\left(-i\frac{\pi}{2}\hat{P}_{23}\left(1+\varepsilon\right)\right) \nonumber \\
&=\hat{P}_{12}\exp\left(-i\frac{\pi}{2}\varepsilon\hat{P}_{12}\right)\hat{P}_{23}\exp\left(-i\frac{\pi}{2}\varepsilon\hat{P}_{23}\right)=\hat{P}_{12}\left(1-i\frac{\pi}{2}\varepsilon\hat{P}_{12}\right)\hat{P}_{23}\left(1-i\frac{\pi}{2}\varepsilon\hat{P}_{23}\right)\nonumber\\
&=\left(\hat{P}_{12}-i\frac{\pi}{2}\varepsilon\mathbf{Id}+\mathrm{O}\left(\varepsilon^{2}\right)\right)\left(\hat{P}_{23}-i\frac{\pi}{2}\varepsilon\mathbf{Id}+\mathrm{O}\left(\varepsilon^{2}\right)\right)\nonumber\\
&=\hat{P}_{12}\hat{P}_{23}-i\frac{\pi}{2}\varepsilon\left(\hat{P}_{12}+\hat{P}_{23}\right)+\mathrm{O}\left(\varepsilon^{2}\right),
\end{align}
where the third line comes from \eqref{properties}, $\hat{P}_{ij}^2=\mathbf{Id}$, and we are ignoring terms above the first order.\\
\\
Exactly the same procedure can be applied for the Hamiltonian in eq.\eqref{secondo}:
\begin{align}\label{diagonale}
&\exp \left(-i\hat{H}T\left(1+\varepsilon\right)\right)=\exp\left(-i\frac{2\pi}{3}\left(\mathbf{Id}+\kappa \hat{P}_{23} \hat{P}_{13}+\kappa^{*} \hat{P}_{13} \hat{P}_{23}\right)(1+\varepsilon)\right)\nonumber\\
&=\hat{P}_{12}\hat{P}_{23}\exp\left(-i\frac{2\pi}{3}\varepsilon\left(\mathbf{Id}+\kappa \hat{P}_{23} \hat{P}_{13}+\kappa^{*} \hat{P}_{13} \hat{P}_{23}\right)\right)\nonumber \\
&=\hat{P}_{12}\hat{P}_{23}\left(1-i\frac{2\pi}{3}\varepsilon\left(\mathbf{Id}+\kappa \hat{P}_{23} \hat{P}_{13}+\kappa^{*} \hat{P}_{13} \hat{P}_{23}\right)+\mathrm{O}\left(\varepsilon^{2}\right)\right)\nonumber \\
&=\hat{P}_{12}\hat{P}_{23}-i\frac{2\pi}{3}\varepsilon\left(\hat{P}_{12}\hat{P}_{23}+\kappa \hat{P}_{12}\hat{P}_{23} \hat{P}_{23} \hat{P}_{13}+\kappa^{*} \hat{P}_{12}\hat{P}_{23} \hat{P}_{13} \hat{P}_{23}\right)+\mathrm{O}\left(\varepsilon^{2}\right).
\end{align} 
This expression can be considerably simplified using \eqref{properties} again and noting that $\hat{P}_{12}\hat{P}_{23} \hat{P}_{13} \hat{P}_{23}$ is the identity permutation, so the final form of the perturbed operator is:
\begin{equation}\label{second}
\hat{P}_{12}\hat{P}_{23}-i\frac{2\pi}{3}\varepsilon\left(\hat{P}_{12}\hat{P}_{23}+\kappa\hat{P}_{12}\hat{P}_{13}+\kappa^*\, \mathbf{Id}\right)+\mathrm{O}\left(\varepsilon^{2}\right)
\end{equation}
The reason why the two results \eqref{first} and \eqref{second} are different lies in the different elements that are perturbed. In fact in the first case we perturbed the single operator, while in the second case the entire Hamiltonian was modified.

Until now we have used approximate results, by always expanding for small $\varepsilon$. Next step will be to consider exact result with arbitrary small perturbation: this should this should come a little closer to realistic situations.

\section{Perturbed eigenstates of the Hamiltonian}
\begin{figure}[h!]
\centering
\includegraphics[scale=0.6]{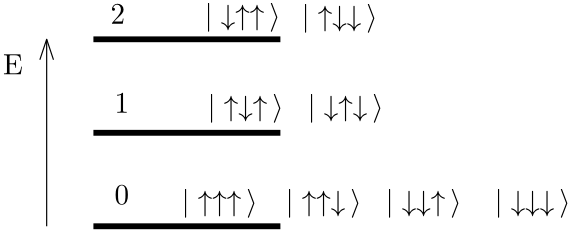}
\vspace{0.5cm}
\caption{Energy levels and respective eigenstates of the three Ising spins system $|s_1\, s_2\, s_3\rangle$ with permutation operator $\hat{P}_{12}\hat{P}_{23}$.}\label{fig:energy}
\end{figure}

In Chapter 2 we briefly talked about the energy levels created by the operator $\hat{P}_{12}\hat{P}_{23}$: the diagonal Hamiltonian \eqref{ham} represents the three degenerate energy levels and the respective eigenstates are presented in relation to \eqref{eigen}. These are fully portrayed for convenience in eqs.\eqref{eigen2} and in fig. \ref{fig:energy}.
\begin{equation*}
\mathbf{v}_1=\begin{pmatrix}
1\\0\\0\\0\\0\\0\\0\\0
\end{pmatrix},\quad
\mathbf{v}_2=\frac{1}{\sqrt{3}}\begin{pmatrix}
0\\1\\
1\\
1\\0\\0\\0\\0
\end{pmatrix},\quad
\mathbf{v}_3=\frac{1}{\sqrt{3}}\begin{pmatrix}
0\\1\\
e^{\frac{-i2\pi}{3}}\\
e^{\frac{i2\pi}{3}}\\0\\0\\0\\0
\end{pmatrix},\quad
\mathbf{v}_4=\frac{1}{\sqrt{3}}\begin{pmatrix}
0\\1\\
e^{\frac{i2\pi}{3}}\\
e^{\frac{-i2\pi}{3}}\\0\\0\\0\\0
\end{pmatrix},
\end{equation*}
\begin{equation}\label{eigen2}
\mathbf{v}_5=\frac{1}{\sqrt{3}}\begin{pmatrix}
0\\0\\0\\0\\1\\
1\\
1\\0
\end{pmatrix},\quad
\mathbf{v}_6=\frac{1}{\sqrt{3}}\begin{pmatrix}
0\\0\\0\\0\\1\\
e^{\frac{-i2\pi}{3}}\\
e^{\frac{i2\pi}{3}}\\0
\end{pmatrix},\quad
\mathbf{v}_7=\frac{1}{\sqrt{3}}\begin{pmatrix}
0\\0\\0\\0\\1\\
e^{\frac{i2\pi}{3}}\\
e^{\frac{-i2\pi}{3}}\\0
\end{pmatrix},\quad
\mathbf{v}_8=\begin{pmatrix}
0\\0\\0\\0\\0\\0\\0\\1
\end{pmatrix}.
\end{equation}

In this context it is be interesting to see how a perturbation can modify the eigenstates or the ontological states \eqref{basis} and if these modifications might include superposition states in the latter case\\
The first approach would be to modify the Hamiltonian as in eq.\eqref{diagonale}:
\begin{align}\label{autoval}
&\hat{U}'=\exp \left(-i\hat{H}T\left(1+\varepsilon\right)\right)\Rightarrow \\
&\Rightarrow\hat{H}'\stackrel{\text{def}}{=}\hat{H}(1+\varepsilon)=\frac{2\pi}{3T}\text{diag}\left\{0,0,1+\varepsilon,2+2\varepsilon,0,1+\varepsilon,2+2\varepsilon,0\right\}\Rightarrow \\
&\Rightarrow\hat{U'}=\text{diag}\left\{1,1,e^{-\frac{i2\pi}{3}}e^{-\frac{i2\pi\varepsilon}{3}},e^{-\frac{i4\pi}{3}}e^{-\frac{i4\pi\varepsilon}{3}},1,e^{-\frac{i2\pi}{3}}e^{-\frac{i2\pi\varepsilon}{3}},e^{-\frac{i4\pi}{3}}e^{-\frac{i4\pi\varepsilon}{3}},1\right\}.
\end{align}
The perturbation is modifying the energy levels in values, but the Hamiltonian remains diagonal, thus the eigenstates don't change in this case.\\ It remains to be seen the effect that $\hat{U}'$ has on the ontological states. We can immediately say that the two states $|\uparrow\uparrow\uparrow\,\rangle$ and $|\downarrow\downarrow\downarrow\,\rangle$ change only by a phase, but otherwise they remain untouched, that's because they are the sole states that are both in the ontological set and in the eigenstates set. Regarding the remaining six states instead, they can be divided in two groups, as done in Chapter 2: three states with two up-spins and three states with two down-spins. Thus it is possible to work with only one of the two groups and the results will be the same for the other group too. This allows us to considerably simplify the ontological set and the eigestates set as:
\begin{equation}\label{ontologici}
\mathbf{s_2}=|\uparrow \uparrow \downarrow \,\rangle=\begin{pmatrix}
1\\0\\0
\end{pmatrix},
\quad\mathbf{s_3}=|\uparrow \downarrow \uparrow\,\rangle=\begin{pmatrix}
0\\1\\0
\end{pmatrix},
\quad\mathbf{s_4}=|\downarrow\uparrow \uparrow\,\rangle=\begin{pmatrix}
0\\0\\1
\end{pmatrix},
\end{equation}
\begin{equation}\label{autostati}
\mathbf{v}_2=\frac{1}{\sqrt{3}}\begin{pmatrix}
1\\
1\\
1
\end{pmatrix},\quad
\mathbf{v}_3=\frac{1}{\sqrt{3}}\begin{pmatrix}
1\\
e^{\frac{-i2\pi}{3}}\\
e^{\frac{i2\pi}{3}}
\end{pmatrix},\quad
\mathbf{v}_4=\frac{1}{\sqrt{3}}\begin{pmatrix}
1\\
e^{\frac{i2\pi}{3}}\\
e^{\frac{-i2\pi}{3}}
\end{pmatrix},
\end{equation}
exactly as already done in sections \ref{sez} and \ref{quelladopo}.\\
In order to apply the perturbed operator $\hat{U}'$ on the ontological states \eqref{ontologici} it is convenient to write each one of them as a linear combination of vectors from \eqref{autostati}, this means that we need to solve system of three linear equations:
\begin{equation}
\mathbf{s_i}=\alpha_i\mathbf{v}_2+\beta_i\mathbf{v}_3+\gamma_i\mathbf{v}_4, \qquad i=2,3,4
\end{equation}
and the resulting coefficients are:
\begin{align}
&\alpha_2=\beta_2=\gamma_2=\frac{\sqrt{3}}{3};\nonumber\\
&\alpha_3=\frac{\sqrt{3}}{3},\quad\beta_3=\frac{i(1-e^{-\frac{i2\pi}{3}})}{3},\quad\gamma_3=-\frac{i(1-e^{\frac{i2\pi}{3}}}{3});\nonumber\\
&\alpha_4=\frac{\sqrt{3}}{3},\quad\beta_4=-\frac{i(1-e^{\frac{i2\pi}{3}})}{3},\quad\gamma_4=\frac{i(1-e^{-\frac{i2\pi}{3}})}{3}.
\end{align}
With these  linear combinations it becomes immediate to apply $\hat{U}'$ on the set \eqref{ontologici}:
\begin{align}\label{caso1}
\hat{U}'|\uparrow\uparrow\downarrow\,\rangle=\hat{U}'\mathbf{s}_2=
\alpha_2\mathbf{v}_2+e^{-\frac{i2\pi}{3}}e^{-\frac{i2\pi\varepsilon}{3}}\beta_2\mathbf{v}_3+e^{-\frac{i4\pi}{3}}e^{-\frac{i4\pi\varepsilon}{3}}\gamma_2\mathbf{v}_4,\nonumber\\
\hat{U}'|\uparrow\downarrow\uparrow\,\rangle=\hat{U}'\mathbf{s}_3=
\alpha_3\mathbf{v}_2+e^{-\frac{i2\pi}{3}}e^{-\frac{i2\pi\varepsilon}{3}}\beta_3\mathbf{v}_3+e^{-\frac{i4\pi}{3}}e^{-\frac{i4\pi\varepsilon}{3}}\gamma_3\mathbf{v}_4,\nonumber \\
\hat{U}'|\downarrow\uparrow\uparrow\,\rangle=\hat{U}'\mathbf{s}_4=
\alpha_4\mathbf{v}_2+e^{-\frac{i2\pi}{3}}e^{-\frac{i2\pi\varepsilon}{3}}\beta_4\mathbf{v}_3+e^{-\frac{i4\pi}{3}}e^{-\frac{i4\pi\varepsilon}{3}}\gamma_4\mathbf{v}_4,
\end{align}
using the eigenvalues from eq.\eqref{autoval}.\\
Going back to the ontological states again, we notice that the results aren't in fact proper ontological states anymore, but instead we obtain superposition states:
\begin{align}
\hat{U}'|\uparrow\uparrow\downarrow\,\rangle&=
|\uparrow\uparrow\downarrow\,\rangle\left[\frac{1}{3}+\frac{1}{3}e^{-\frac{i2\pi}{3}}e^{-\frac{i2\pi\varepsilon}{3}}+\frac{1}{3}e^{-\frac{i4\pi}{3}}e^{-\frac{i4\pi\varepsilon}{3}}\right]+
\nonumber \\
&+|\uparrow\downarrow\uparrow\,\rangle\left[\frac{1}{3}+\frac{e^{-\frac{i2\pi}{3}}}{3}e^{-\frac{i2\pi}{3}}e^{-\frac{i2\pi\varepsilon}{3}}+\frac{e^{\frac{i2\pi}{3}}}{3}e^{-\frac{i4\pi}{3}}e^{-\frac{i4\pi\varepsilon}{3}}\right]+
\nonumber \\
&+|\uparrow\uparrow\downarrow\,\rangle\left[\frac{1}{3}+\frac{e^{\frac{i2\pi}{3}}}{3}e^{-\frac{i2\pi}{3}}e^{-\frac{i2\pi\varepsilon}{3}}+\frac{e^{-\frac{i2\pi}{3}}}{3}e^{-\frac{i4\pi}{3}}e^{-\frac{i4\pi\varepsilon}{3}}\right]
=\nonumber\\
&=
|\uparrow\uparrow\downarrow\,\rangle\frac{1}{3}\left[1-e^{-\frac{i2\pi\varepsilon}{3}}\left(\left(\frac{1}{2}+\frac{i\sqrt{3}}{2}\right)+\left(\frac{1}{2}-\frac{i\sqrt{3}}{2}\right)e^{-\frac{i2\pi\varepsilon}{3}}\right)\right]+
\nonumber \\
&+|\uparrow\downarrow\uparrow\,\rangle\frac{1}{3}\left[1-e^{-\frac{i  2 \pi\varepsilon }{3}}\left(\left(\frac{1}{2}-\frac{i\sqrt{3}}{2}\right)+\left(\frac{1}{2}+\frac{i\sqrt{3}}{2}\right)e^{-\frac{i 2  \pi\varepsilon }{ 3}}\right)\right]+
\nonumber \\
&+|\downarrow\uparrow\uparrow\,\rangle\frac{1}{3}\left[1+e^{-\frac{i 2 \pi\varepsilon}{3}}\left(1+e^{-\frac{i2\pi\varepsilon}{3} }\right)\right];
\end{align}
\begin{align}
\hat{U}'|\uparrow\downarrow\uparrow\,\rangle&=
|\uparrow\uparrow\downarrow\,\rangle\left[\frac{1}{3}+\frac{i\left(1-e^{-\frac{i2\pi}{3}}\right)}{3\sqrt{3}}e^{-\frac{i2\pi}{3}}e^{-\frac{i2\pi\varepsilon}{3}}-\frac{i\left(1-e^{\frac{i2\pi}{3}}\right)}{3\sqrt{3}}e^{-\frac{i4\pi}{3}}e^{-\frac{i4\pi\varepsilon}{3}}\right]+
\nonumber \\
&+|\uparrow\downarrow\uparrow\,\rangle\left[\frac{1}{3}+i\left(1-e^{-\frac{i2\pi}{3}}\right)\frac{e^{-\frac{i2\pi}{3}}}{3\sqrt{3}}e^{-\frac{i2\pi}{3}}e^{-\frac{i2\pi\varepsilon}{3}}-i\left(1-e^{\frac{i2\pi}{3}}\right)\frac{e^{\frac{i2\pi}{3}}}{3\sqrt{3}}e^{-\frac{i4\pi}{3}}e^{-\frac{i4\pi\varepsilon}{3}}\right]+
\nonumber \\
&+|\downarrow\uparrow\uparrow\,\rangle\left[\frac{1}{3}+i\left(1-e^{-\frac{i2\pi}{3}}\right)\frac{e^{\frac{i2\pi}{3}}}{3\sqrt{3}}e^{-\frac{i2\pi}{3}}e^{-\frac{i2\pi\varepsilon}{3}}-i\left(1-e^{\frac{i2\pi}{3}}\right)\frac{e^{-\frac{i2\pi}{3}}}{3\sqrt{3}}e^{-\frac{i4\pi}{3}}e^{-\frac{i4\pi\varepsilon}{3}}\right]
=\nonumber \\
&=
|\uparrow\uparrow\downarrow\,\rangle\frac{1}{3}\left[1+ e^{-\frac{i 2 \pi \varepsilon}{3}}\left(1+e^{-\frac{i 2 \pi \varepsilon}{3}}\right)\right]+
\nonumber \\
&+|\uparrow\downarrow\uparrow\,\rangle\frac{1}{3}\left[1-e^{-\frac{i2\pi\varepsilon}{3}}\left(\left(\frac{1}{2}+\frac{i3}{2\sqrt{3}}\right)+\left(\frac{1}{2}-\frac{i3}{2\sqrt{3}}\right)e^{-\frac{i2\pi\varepsilon}{3}}\right)\right]+
\nonumber \\
&+|\downarrow\uparrow\uparrow\,\rangle\frac{1}{3}\left[1-e^{-\frac{i2\pi\varepsilon}{3}}\left(\left(\frac{1}{2}-\frac{i3}{2\sqrt{3}}\right)+\left(\frac{1}{2}+\frac{i3}{2\sqrt{3}}\right)e^{-\frac{i2\pi\varepsilon}{3}}\right)\right];
\end{align}
\begin{align}
\hat{U}'|\downarrow\uparrow\uparrow\,\rangle&=
|\uparrow\uparrow\downarrow\,\rangle\left[\frac{1}{3}-\frac{i\left(1-e^{\frac{i2\pi}{3}}\right)}{3\sqrt{3}}e^{-\frac{i2\pi}{3}}e^{-\frac{i2\pi\varepsilon}{3}}+\frac{i\left(1-e^{-\frac{i2\pi}{3}}\right)}{3\sqrt{3}}e^{-\frac{i4\pi}{3}}e^{-\frac{i4\pi\varepsilon}{3}}\right]+
\nonumber \\
&+|\uparrow\downarrow\uparrow\,\rangle\left[\frac{1}{3}-i\left(1-e^{\frac{i2\pi}{3}}\right)\frac{e^{-\frac{i2\pi}{3}}}{3\sqrt{3}}e^{-\frac{i2\pi}{3}}e^{-\frac{i2\pi\varepsilon}{3}}+i\left(1-e^{-\frac{i2\pi}{3}}\right)\frac{e^{\frac{i2\pi}{3}}}{3\sqrt{3}}e^{-\frac{i4\pi}{3}}e^{-\frac{i4\pi\varepsilon}{3}}\right]+
\nonumber \\
&+|\downarrow\uparrow\uparrow\,\rangle\left[\frac{1}{3}-i\left(1-e^{\frac{i2\pi}{3}}\right)\frac{e^{\frac{i2\pi}{3}}}{3\sqrt{3}}e^{-\frac{i2\pi}{3}}e^{-\frac{i2\pi\varepsilon}{3}}+i\left(1-e^{-\frac{i2\pi}{3}}\right)\frac{e^{-\frac{i2\pi}{3}}}{3\sqrt{3}}e^{-\frac{i4\pi}{3}}e^{-\frac{i4\pi\varepsilon}{3}}\right]
=\nonumber \\
&=
|\uparrow\uparrow\downarrow\,\rangle\frac{1}{3}\left[1-e^{-\frac{i2\pi\varepsilon}{3}}\left(\left(\frac{1}{2}-\frac{i3}{2\sqrt{3}}\right)+\left(\frac{1}{2}+\frac{i3}{2\sqrt{3}}\right)e^{-\frac{i2\pi\varepsilon}{3}}\right)\right]
+\nonumber \\
&+
|\uparrow\downarrow\uparrow\,\rangle\frac{1}{3}\left[1+ e^{-\frac{i 2 \pi \varepsilon}{3}}\left(1+e^{-\frac{i 2 \pi \varepsilon}{3}}\right)\right]+
\nonumber \\
&+
|\downarrow\uparrow\uparrow\,\rangle\frac{1}{3}\left[1-e^{-\frac{i2\pi\varepsilon}{3}}\left(\left(\frac{1}{2}+\frac{i3}{2\sqrt{3}}\right)+\left(\frac{1}{2}-\frac{i3}{2\sqrt{3}}\right)e^{-\frac{i2\pi\varepsilon}{3}}\right)\right].
\end{align}

All the above equations give the correct results for $\hat{U}'\mapsto \hat{U}$ as $\varepsilon\mapsto0$.\\Exactly the same result can be obtained using the set of ontological states with two down-spins: it is enough to apply the following substitutions:
\begin{equation}
|\uparrow\uparrow\downarrow\,\rangle\mapsto
|\downarrow\downarrow\uparrow\,\rangle,\quad
|\uparrow\downarrow\uparrow\,\rangle\mapsto
|\downarrow\uparrow\downarrow\,\rangle,\quad
|\downarrow\uparrow\uparrow\,\rangle\mapsto
|\uparrow\downarrow\downarrow\,\rangle.
\end{equation}

\subsection{A more extensive example}
Another possible case is a generic diagonal perturbation $\hat{A}=\text{diag}\{c_1,c_2,c_3,c_4,c_5,c_6,c_7,c_8\}$, $c_i \in \mathbb{C}$ so that $\hat{H}'=\hat{H}+\hat{A}$.\\
The correction $\hat{A}$ can be interpreted as a more general case than \eqref{autoval}, which could give a more realistic description of the inaccuracies of the eigenvalues because of the high number of parameters which cover all the possible degrees of freedom.

Of course, $\hat{A}$ creates superpositions too if applied on the ontological states: without going into details, the eigenstates of the new permutation operator become:
\begin{equation}
\hat{U'}=\text{diag}\left\{e^{ic_1 T},e^{ic_2 T},e^{-\frac{i2\pi}{3}}e^{ic_3 T},e^{-\frac{i4\pi}{3}}e^{ic_4 T},e^{ic_5 T},e^{-\frac{i2\pi}{3}}e^{ic_6 T},e^{-\frac{i4\pi}{3}}e^{ic_7 T},e^{ic_8 T}\right\}.
\end{equation}
Using again the ontological set \eqref{ontologici} and the eigenstates (which again remain unchanged) \eqref{autostati}, the trasformation on the ontological states is:
\begin{align}\label{banale}
\hat{U}'|\uparrow\uparrow\downarrow\,\rangle=\hat{U}'\mathbf{s}_2=
e^{ic_2 T}\alpha_2\mathbf{v}_2+e^{-\frac{i2\pi}{3}}e^{ic_3 T}\beta_2\mathbf{v}_3+e^{-\frac{i4\pi}{3}}e^{ic_4 T}\gamma_2\mathbf{v}_4,
\nonumber\\
\hat{U}'|\uparrow\downarrow\uparrow\,\rangle=\hat{U}'\mathbf{s}_3=
e^{ic_2 T}\alpha_3\mathbf{v}_2+e^{-\frac{i2\pi}{3}}e^{ic_3 T}\beta_3\mathbf{v}_3+e^{-\frac{i4\pi}{3}}e^{ic_4 T}\gamma_3\mathbf{v}_4,
\nonumber \\
\hat{U}'|\downarrow\uparrow\uparrow\,\rangle=\hat{U}'\mathbf{s}_4=
e^{ic_2 T}\alpha_4\mathbf{v}_2+e^{-\frac{i2\pi}{3}}ee^{ic_3 T}\beta_4\mathbf{v}_3+e^{-\frac{i4\pi}{3}}e^{ic_4 T}\gamma_4\mathbf{v}_4.
\end{align}
Keeping in mind the results following from eq.\eqref{caso1}, it is trivial to prove that eq.\eqref{banale} results in superpositions of ontological states again. This would not be true if it happens that $c_2= c_3= c_4$ or $c_5=c_6=c_7$, but it is an exceptional event that would not reflect a real measurement: in fact it is unlikely, if not impossible, to get the exact same error measuring three different energy eigenvalues. So in order to produce the most general simulation of what could happen during an experiment, we imagine that for each eigenstate a researcher would measure its energy level along with its unique error.

These are few examples of how some kind of perturbations, caused by the finite precision of a measuring instrument, can trigger an apparent quantum behaviour on a classical system.\\
There exist other ways to modify the evolution operator or the Hamiltonian, for example, by adding to $\hat{H}$ a non diagonal or non Hermitian matrix, or acting on eq.\eqref{quantistica} by slightly modifying a coefficient for some Pauli matrix.

\chaptermark{Conclusions}
\chapter*{Conclusions}
In this thesis we addressed the problem of the non classical features of Quantum Mechanics (QM), focusing on the superposition of states. We considered an interpretation of QM that is an alternative to the Copenhagen one, the Cellular Automata Interpretation \cite{Hooft2014}, which proposes an alternative explanation of the quantum paradoxes and imposes a continuity of the classical behaviour from the macroscopic to the microscopic world.\\
Cellular Automata (CA) are discrete objects that evolve following classical and deterministic laws, through discrete space and time. This is a model that finds applications in many scientific contexts, in particular in \cite{Hooft2014} it is used to describe the states of particles at the atomic/sub-atomic level.\\
The fundamental idea of the CA interpretation is that microscopic particles evolve through classical states following the CA model, these states are called Ontological States; Quantum States are instead the result of the even small imprecisions in measurements.\\
These are the main principles on which this thesis is based: our aim was to prove that, in a certain classical system, imprecisely known values or parameters cause superposition of states.\\ In order to do that, we used the system introduced in \cite{Elze2020} which consists of a triplet of coupled Ising spins. Ising spins are particularly suitable for this purpose since they are classical objects that have a quantum counterpart, qubits: in fact, in \cite{Elze2020}, while building the time evolution operator, said operator has been rewritten using Pauli matrices to fit it in a Hilbert space and to make it possible to use said operator on qubits. The time evolution operator is a permutation that exchanges the states of the three spins: it strictly obeys to the CA fundamentals, being discrete and deterministic.\\
Following closely the dynamics outlined in \cite{Elze2020} we applied a series of perturbations to the system, to verify whether superposition states would arise. The first attempt has been to adjust the time evolution operator with a small correction of order $\varepsilon$, both on the definition of the permutation operator and on the Hamiltonian, using in this case a finite Baker-Hausdorff-Campbell series introduced in \cite{Elze2020}. In both cases superposition states arise as a result of a Taylor expansion in $\varepsilon$.\\
We then considered a more realistic case, introducing imprecisions directly on the eigenvalues of the Hamiltonian without using any kind of approximation. Firstly, we imposed the same correction of order $\varepsilon$ on all eigenstates, finally we added to the diagonalized Hamiltonian a diagonal matrix that would simulate a different degree of imprecision for each eigenvalue. The latter case would reflect closer what can happen when measuring the energy levels of a system, whule the ontological degree of freedom are still unknown. This would be in fact the most general case, hypothesizing that, during an experiment, a researcher would get for each eigenstate a different error for the measured energy level, even amongst the degenerate ones.
In both cases, when applying the perturbed time evolution operator on the ontological states, these will result in quantum superposition states.
In a future work there could be other ways to implement different kinds of perturbations, for example using non Hermitian matrices. Nevertheless, with simple perturbations, we have successfully triggered quantum features in a classical system; this is a first step in considering quantum behaviour not as intrinsic of microscopic systems anymore, but as a result of the unavoidable finite precision of measuring instruments.

\addcontentsline{toc}{chapter}{Conclusions}
\bibliography{bibliografia}

\begin{thebibliography}{10}

\bibitem{Hooft2014}
G.~'t~Hooft, {\em The {Cellular Automaton Interpretation of Quantum
  Mechanics}}.
\newblock Vol. 185 of \textit{Fundamental Theories of Physics}, Springer
  Nature, 2016.

\bibitem{Elze2020}
H.-T. Elze, ``{Qubit exchange interactions from permutations of classical
  bits},'' {\em Int. J. of Quantum Information}, vol.~17, no.~8, p.~1941003,
  2019.

\bibitem{Ulam1952}
S.~Ulam, ``Random processes and transformations,'' {\em Proc. Int. Congress on
  Mathematics}, vol.~2, pp.~264--275, 1952.

\bibitem{Neumann1951}
J.~Von~Neumann {\em et~al.}, ``The general and logical theory of automata,''
  {\em Cerebral Mechanisms in Behavior—The Hixon Symposium}, pp.~1--31, 1951.

\bibitem{Gardener1970}
M.~Gardener, ``The fantastic combinations of {John Conway's} new solitaire game
  "life",'' {\em Sci. Am.}, vol.~223(4), 1970.

\bibitem{Gardener1971}
M.~Gardener, ``On cellular automata, self-reproduction, the {Garden Eden} and
  the game “life”,'' {\em Sci. Am.}, vol.~224(2), 1971.

\bibitem{Wolfram1983}
S.~Wolfram, ``Statistical mechanics of cellular automata,'' {\em Rev. Mod.
  Phys.}, vol.~55, pp.~601--644, 1983.

\bibitem{Wolfram2002}
S.~Wolfram, {\em A new kind of science}.
\newblock Wolfram Media, Champaign, 2002.

\bibitem{Russell1945}
B.~Russell, {\em History of western philosophy.}
\newblock Saggistica TES 8va edizione, 1945.

\bibitem{Severino1996}
E.~Severino, {\em {La filosofia dai Greci al nostro tempo. La filosofia
  moderna}}.
\newblock BUR Saggi, 7ima edizione, 1996.

\bibitem{Elze2002}
H.-T. Elze and O.~Schipper, ``{Time without time: A stochastic clock model},''
  {\em Phys. Rev. D}, vol.~66, no.~4, p.~2, 2002.

\bibitem{Elze2003}
H.-T. Elze, ``Emergent discrete time and quantization: Relativistic particle
  with extra-dimensions,'' {\em Phys. Lett. A}, vol.~310, no.~2-3,
  pp.~110--118, 2003.

\bibitem{Elze20202}
H.-T. Elze, ``{A Baker-Campbell-Hausdorff formula for the logarithm of
  permutations},'' {\em Int. J. of Geometric Methods in Modern Physics},
  vol.~17, no.~4, 2020.

\bibitem{Shannon1949}
C.~E. Shannon, ``{Communication in the Presence of Noise},'' {\em Proceedings
  of the IRE}, vol.~37, no.~1, pp.~10--21, 1949.

\bibitem{Elze2014}
H.-T. Elze, ``Action principle for cellular automata and the linearity of
  quantum mechanics,'' {\em Phys. Rev. A}, vol.~89, no.~1, p.~012111, 2014.

\end{thebibliography}
\addcontentsline{toc}{chapter}{Bibliography}

\appendix

\chaptermark{Appendix}

\chapter{The Sampling Theorem}

Here we would like to clarify some formal aspects about the use of a discrete time in the Cellular Automata interpretation.\\Since there is supposed to be a more or less direct link between CA systems and the corresponding Quantum Mechanics model, to achieve that the Sampling Theorem \cite{Shannon1949} is used: it creates a mapping between the discrete time of CA and the quantum continuous time \cite{Elze2014}. In order to do that, let's first define a \textit{band-limited function} $f(t)$:
\begin{defn}
Given a function $f(t)$ and be $\mathcal{F}(f)$ its Fourier transform:
\begin{equation*}
f(t)\quad \text{is band limited if}\quad f \in L^{2},\quad \mathcal{F}(f) \in \left[-\omega_{max}, \omega_{max}\right],\quad \omega_{max}\equiv\,\,\text{bandwidth}.
\end{equation*}
\end{defn}
\begin{theorem*}[Sampling Theorem]
Given $f(t)$ an integrable, band-limited function, 
\begin{equation*}
f(t)=\frac{1}{2 \pi} \int_{-\omega_{max}}^{\omega_{max}} e^{i \omega t} f(\omega) d \omega,
\end{equation*}
and a set of values of the function $\{f(t_{n})\}$ at equidistantly spaced times $t_n$ so that $\left|t_{n}-t_{n-1}\right|=\frac{\pi}{\omega_{m a x}}$. Then it holds that $f(t)$ can be rewritten as:
\begin{equation}\label{function}
f(t)=\sum_{n} f\left(t_{n}\right) \frac{\sin \left[\omega_{\max }\left(t-t_{n}\right)\right]}{\omega_{\max }\left(t-t_{n}\right)}.
\end{equation} 
\end{theorem*}
This theorem is quite useful in our context, because as long as eq.\eqref{function} is exact, it is possible to use the succession of discrete values $\{t_{n}\}$ to create a discrete time $n$ for CA, with the two times related by a proportionality constant $l$, $t_{n}=nl$ (note that in this thesis the notation $T$ has been used in place of $l$). The number $l$ is the fundamental scale of this discrete parametrization of time, and it is such that $\omega_{max}=\frac{\pi}{l}$.

\end{document}